\input harvmac

%%%%%%%%%%%%%%%%%%%%%%%%%%%%%%%%%%%%%%%%%%%%%%%%%%%%%%%%%%%%%%%%%%%
%%%  modify title page
%%%%%%%%%%%%%%%%%%%%%%%%%%%%%%%%%%%%%%%%%%%%%%%%%%%%%%%%%%%%%%%%%%%
\def\Title#1#2{\rightline{#1}\ifx\answ\bigans\nopagenumbers\pageno0
\vskip0.5in
\else\pageno1\vskip.5in\fi \centerline{\titlefont #2}\vskip .3in}

\font\caps=cmcsc10
%\def\listrefs{\footatend\bigskip\bigskip\immediate\closeout\rfile
%\writestoppt \baselineskip =13pt\centerline{{\secfont References}}
%\bigskip{\frenchspacing\parindent =20pt \escapechar +'
%\input\jobname.refs \vfill\eject}\nonfrenchspacing} 
%%%%%%%%%%%%%%%%%%%%%%%%%%%%%%%%%%%%%%%%%%%%%%%%%%%%%%%%%%%%%%%%%%%%%%%%%%%%

\noblackbox
\parskip=1.5mm
%\def\semi{;~}

%%%%%%%%%%%%%%%%%%%%%%%%%%%%%%%%%%%%%%%%%%%%%%%%%%%%%%%%%%%%%%%%%%%%%
  
\def\npb#1#2#3{{\it Nucl. Phys.} {\bf B#1} (#2) #3 }
\def\plb#1#2#3{{\it Phys. Lett.} {\bf B#1} (#2) #3 }
\def\prd#1#2#3{{\it Phys. Rev. } {\bf D#1} (#2) #3 }
\def\prl#1#2#3{{\it Phys. Rev. Lett.} {\bf #1} (#2) #3 }
\def\mpla#1#2#3{{\it Mod. Phys. Lett.} {\bf A#1} (#2) #3 }

\def\cmp#1#2#3{{\it Commun. Math. Phys.} {\bf #1} (#2) #3 }

\def\bb#1{{\tt hep-th/#1}}

%%%%%%%%%%%%%%%%%%%%%%%%%%%%%%%%%%%%%%%%%%%%%%%%%%%%%%%%%%%%%%%%%%%%%
%%%%%%%%%%%%%%%%%%%%    some definitions    %%%%%%%%%%%%%%%%%%%%%%%%%
%%%%%%%%%%%%%%%%%%%%%%%%%%%%%%%%%%%%%%%%%%%%%%%%%%%%%%%%%%%%%%%%%%%%%

           \def\CO{{\cal O}}

\def\CM{{\cal M}} 
\def\CN{{\cal N}}

%%%%%%%%%%%%%%%%%%%%%%%%%%%%%%%%%%%%%%%%%%%%%%%%%%%%%%%%%%%%%%%%%%%%%

\def\dj{\hbox{d\kern-0.347em \vrule width 0.3em height 1.252ex depth
-1.21ex \kern 0.051em}}

\def\half{{1\over 2}\,}

\def\ket{\rangle}
\def\bra{\langle}

\def\balpha{\overline \alpha}
\def\bS{\overline S}
\def\bQ{\overline Q}
\def\pt{\partial}

%%%%%%%%%%%%%%%%%%%%%%%%%%%%%%%%%%%%%%%%%%%%%%%%%%%%%%%%%%%%%%%%%%%%%%
%%%%%%%%%%%%%%%%%%%%%%%        references         %%%%%%%%%%%%%%%%%%%%%%
%%%%%%%%%%%%%%%%%%%%%%%%%%%%%%%%%%%%%%%%%%%%%%%%%%%%%%%%%%%%%%%%%%%%%%%dsky
\lref\rgreenf{M.B. Green, \plb {329}{1994}{435,} \bb{9403040.}}
\lref\rpolsem{J. Dai, R.G. Leigh and J. Polchinski, \mpla{4}
 {1989}{2073.}}
\lref\rpolrr{J. Polchinski, \prl{75}{1995}{4728,} \bb{9510017.}}
\lref\rgreent{M.B. Green, \plb {282}{1992}{380,} \bb{9210054.}}
\lref\rgreenb{M.B. Green, \plb {266}{1991}{325.}} 
\lref\ranalogo{J. Polchinski, \cmp{104}{1986}{37 \semi}
K.H. O'Brien and C.I. Tan, \prd{36}{1987}{1184\semi} 
B. McClain and B.D.B. Roth, \cmp{111}{1987}{539\semi}
E. Alvarez and M.A.R. Osorio, \prd{40}{1989}{1150.}}
\lref\raw{J. Atick and E. Witten, \npb{310}{1988}{291.}}
\lref\rpoint{E.F. Corrigan and D.B. Fairlie, \npb{91}{1975}{527\semi} 
M.B. Green, \plb{69}{1976}{89\semi}
I.R. Klebanov and L. Thorlacius, \plb {371}{1996}{51,}  \bb{9510200\semi}
J.L.F. Barb\'on, \plb {382}{1996}{60,} \bb{9601098.}}  
\lref\rcdg{C.G. Callan, R. Dashen and D.J. Gross, \prd {17}{1978}{2717.}}
\lref\raffleck{I. Affleck, \npb {191}{1981}{429.}}
\lref\rbsus{T. Banks and L. Susskind, RU-95-87,  \bb{9511194.}}
\lref\rpolcom{J. Polchinski, \prd {50}{1994}{6041,} \bb{9407031.}}
\lref\rmolive{C. Montonen and D. Olive, \plb {72}{1977}{117.}}
\lref\rgpy{D.J. Gross, R. Pisarski and L. Yaffe, {\it Rev. Mod. Phys.} 
{\bf 53} (1981) 43.}     
\lref\rahw{I. Affleck, J. Harvey and E. Witten, \npb {206}{1982}{413.}}
\lref\rtd{P. Horava, \plb {231}{1989}{251\semi}
J. Polchinski, S. Chaudhuri and C.V. Johnson, NSF-ITP-96-003, 
\bb{9602052\semi}
E. Alvarez, J.L.F. Barb\'on and J. Borlaf, PUPT-96-1601, \bb{9603089.}}
\lref\rgreengut{M.B. Green and M. Gutperle, \npb {476}{1996}{484,}
 \bb{9604091.}}
\lref\rgreengutt{M.B. Green and M. Gutperle, DAMTP-96-110, \bb{9612127.}}
\lref\rvafa{H. Ooguri and C. Vafa, HUTP-96-A036, \bb{9608079.}}
\lref\rmottola{E. Mottola, \prd {17}{1978}{1103.}} 
\lref\rbbs{K. Becker, M. Becker and A. Strominger, \npb {456}{1995}{130,}
 \bb{9507158.}}
\lref\ritep{V.A. Novikov, M.A. Shifman, A.I. Vainshtein, V.B. Voloshin 
and V.I. Zakharov, \npb {229}{1983}{394.}}
\lref\rpoly{A.M. Polyakov, \npb {120}{1977}{429.}}
\lref\rsch{J.H. Schwarz, \npb {226}{1983}{269.}}
\lref\rgs{M.B. Green and J.H. Schwarz, \plb {122}{1983}{143.}}
\lref\rma{M.A. V\'azquez-Mozo, IASSNS-HEP-96-73, \bb{9607052.}}
\lref\rggp{G.W. Gibbons, M.B. Green and M.J. Perry, \plb {370}{1996}{37,}
\bb{9511080.}}
\lref\rwit{E. Witten, \npb {474}{1996}{343,} \bb{9604030.}}
\lref\rgreengas{M.B. Green, \plb {354}{1995}{271,} \bb{9504108.}}
\lref\rutgers{M.R. Douglas, D. Kabat, P. Pouliot and S. Shenker, 
RU-96-62, \bb{9608024.}}
\lref\rwittb{E. Witten, \npb {460}{1996}{335,} \bb{9510135.}}
\lref\rfint{J.L.F. Barb\'on and M.A. V\'azquez-Mozo, CERN-TH/96-361,
 IASSNS-96/127, 
\bb{9701142.}}  
\lref\rhama{K. Hamada, KEK-TH-504, \bb{9612234.}}
\lref\rgpeff{M.B. Green and M. Gutperle, DAMTP/96-104, \bb{9701093.}}   
%%%%%%%%TEXT%%%%%%%%%%%%%%%%%%%%%%%%%%%%%%%%%%%%%%%%%%%%%%%%%%%%%%%%%
%%%%%%%%%%%%%%%%%%%%%%%%%%%%%%%%%%%%%%%%%%%%%%%%%%%%%%%%%%%%%%%%%%%%%B
%%%%%%%%%%%%%%%%%%          title page       %%%%%%%%%%%%%%%%%%%%%%%%%
%%%%%%%%%%%%%%%%%%%%%%%%%%%%%%%%%%%%%%%%%%%%%%%%%%%%%%%%%%%%%%%%%%%%%%

\line{\hfill CERN-TH/96-360}
\line{\hfill {\tt hep-th/9701075}}
\vskip 0.5cm

\Title{\vbox{\baselineskip 12pt\hbox{}
 }}
{\vbox {\centerline{Fermion Exchange between D-instantons  }
}}

\centerline{$\quad$ {\caps J. L. F. Barb\'on
 }}
\smallskip

\centerline{{\sl Theory Division, CERN}}
\centerline{{\sl 1211 Geneva 23. Switzerland}}
\centerline{{\tt barbon@mail.cern.ch}}

 \vskip 1.0in

We define fermionic collective coordinates for type-IIB
 Dirichlet instantons
and discuss some effects of the associated fermionic zero modes
within the dilute gas framework. We show that the standard rules
for clustering of zero modes in the dilute limit, and the 
fermion-exchange interactions follow from world-sheet Ward
identities.   Fermion exchange 
is strongly attractive at string-scale distances, which makes the  
 short-distance Hagedorn
singularity between instantons and anti-instantons  even 
stronger.           

%%%%%%%%%%%%%%%%%%%%%%%%%%%%%%%%%%%%%%%%%%%%%%%%%%%%%%%%%%%%%%%%%%%%%%

\Date{December 1996 }
%\draft

%%%%%%%%%%%%%%%%%%%%%%%%%%%%%%%%%%%%%%%%%%%%%%%%%%%%%%%%%%%%%%%%%%%%%%%%%%
%%%%%%%%%%%%                text begins                        %%%%%%%%%%%
%%%%%%%%%%%%%%%%%%%%%%%%%%%%%%%%%%%%%%%%%%%%%%%%%%%%%%%%%%%%%%%%%%%%%%%%%%

\newsec{Introduction}
The Dirichlet-brane construction of Ramond--Ramond solitons 
\refs\rpolsem, \rpolrr\  allows a
rather explicit characterization of the weak coupling quantization of
such objects. The collective dynamics of a $p$-brane consists of a
world-volume interacting 
 theory of open strings propagating in $p+1$ dimensions, and  the
description of the interactions with the bulk dynamics is equally
explicit in terms of the standard coupling between open and closed
strings.                                 
A special role is played by the Dirichlet instanton of the type-IIB 
theory, the case $p=-1$
localized in all space-time directions. The ``world-volume" is a point,
and the collective dynamics is given by a finite-dimensional integral
over the multi-instanton moduli space, the integrand being determined by
a ``zero-dimensional" open string theory. Besides the simpler
collective dynamics, D-instantons exhibit a number of special
properties, as compared to the rest of the D-branes. It is well known
that perturbative string amplitudes in the instanton background are
power-behaved at high energy
 \refs\rpoint, thus introducing ``field theoretical"
features in string theory. The kind of logarithmic divergences
responsible for the world-volume dynamics in the higher branes (via a
Fischler--Susskind mechanism), are cancelled in this case by the
instanton gas combinatorics, 
 which also ensures the standard clustering
properties \refs\rgreengas, \refs\rpolcom.  

The analogue of the static interaction potential for higher branes is
the interaction action of D-instantons. 
 The leading-order contribution in string
perturbation theory is the cylinder diagram with two Dirichlet
boundaries. In terms of massless background fields \rggp\ it
corresponds to the purely classical overlapping interaction of the
long-distance tails of the instanton fields, as in 
\refs\rcdg. In the closed-string
channel it can be interpreted as an off-shell propagator
between Dirichlet boundary states
\eqn\cilindro
{\Gamma (x,y) \sim \sum_s C_s \,  \left\bra B_x, s\,      
 \left|{1\over \Delta_s}\,P_{GSO}\, \right|B'_y, s \right\ket 
,}
where $C_s$ are appropriate phases for the coherent sum of spin
structures in the GSO-projected 
closed-string channel, and $\Delta_s = {\alpha' \over
2} p^2 + N + {\overline N} -a_s - {\bar a}_s $ is the world-sheet
Hamiltonian for type II strings.           This static interaction
vanishes for two instantons or two anti-instantons as a result of an
unbroken supersymmetry in the open string channel (where we have a
one-loop vacuum diagram of open strings with fixed endpoints, see 
\rgreenf\ for explicit expressions of \cilindro).
In the closed-string channel, this
 is just the well known  ``zero-force" property of
BPS saturated configurations \refs\rmolive, \rgreenf,  \refs\rpolrr.
At the massless level, the cancellation results from the balance
between the Coulomb interactions $\Gamma (x,y)\sim |x-y|^{-8}$,
 due to the 
attractive NS--NS dilaton exchange, and the
repulsive R--R ``axion" exchange. On the other hand, the
R--R scalar exchange becomes attractive between instantons and
anti-instantons, a configuration that breaks all the supersymmetries:
\eqn\iain{\Gamma_{+-} (x-y) = -(2\pi)^4 \int_{0}^{\infty} {ds\over s^5}
\, e^{-{(x-y)^2 \over 2\alpha' s}} \, \prod_{m=1}^{\infty} \left(
{1+q^{2m} \over 1-q^{2m}} \right)^8 ,} where $q=e^{-s}$ is the modular
parameter of the cylindrical world-sheet. 
At short distances, this instanton/anti-instanton (I--A) interaction
suffers from a Hagedorn singularity: each massive closed string
contributing to \iain\ is suppressed by a factor $e^{-M|x-y|}$, and
we can estimate the propagator as $\sum_M \rho_D (M) e^{-M|x-y|}$,
where $\rho_D (M)$ is the level density of closed string states
coupling to the D-instanton. This quantity grows exponentially with
half the rate of the  total density of states $\rho_D (M) \sim
e^{{\beta_{\rm Hag} \over 2}M}$, which   results in  an  
instability at half the Hagedorn distance \refs\rgreenb, \refs\rgreenf.  
Since \iain\ diverges at the ultraviolet endpoint $s\sim 0$, it is
convenient to perform a modular transformation $t = 2\pi^2 /s$ to the
open-string channel, which is also useful to define the absolute
normalization of the amplitude:
\eqn\iaino{\eqalign{\Gamma_{+-} (x-y) =& -2\cdot {1\over 4} \int_{0}^{
\infty} {dt\over t} \Big[ {\rm Tr}_{NS} \left( 1+(-1)^F \right) e^{-t
\Delta_{NS}} - {\rm Tr}_R \left( 1+(-1)^F \right) e^{-t\Delta_R } \Big]
\cr  
=& -\int_{0}^{\infty} {dt\over t} \, e^{-t \left( {(x-y)^2 \over 4 \pi^2
\alpha'} - \half \right)} \, \prod_{m=1}^{\infty} \left( {1-w^{m-1/2} 
\over 1- w^m }\right)^8 }}
where $w= e^{-t}$ and the overall factor of two is due to the oriented
character of the type-IIB strings \refs\rpolrr. In what is by now
a well understood rule \refs\rutgers, the open-string channel is the
appropriate one to discuss short-distance dynamics of D-branes. Indeed,
the massless states producing the $t\sim \infty$ singularities in \iaino\
are associated
 to a vanishing Dirichlet open-string world-sheet Hamiltonian
in the Neveu-Schwarz sector $\Delta_{NS} = {(x-y)\over 4\pi^2 \alpha'} 
+N-\half =0$, as a result of the balance between the Casimir and    
stretching energy of the Dirichlet open string.  
From \iaino\ it follows that the dominant singularity occurs at
half the Hagedorn distance of the type-IIB theory $\beta_{\rm Hag} =
\pi\sqrt{8\alpha'}$, and it is logarithmically attractive: $\Gamma (x,y)
\sim {\rm log} (|x-y| -\beta_{\rm Hag} /2)$. 

 An analogous singularity exists for
higher $p$-branes, for which  some  derivative of the static force
 diverges  \refs\rbsus. These singularities indicate that I--A
configurations are not appropriate varibles at string-length
distances. Indeed, on general grounds, I--A configurations annihilate
one another and cannot be distinguished from perturbative fluctuations
in the coincidence limit, and their treatment is always very ambiguous
at short distances. It is nevertheless surprising that a stringy
instability poses a clear-cut limit to the I--A parametrization.

In the present paper we
 consider some  effects on the I--A interaction due to 
 supersymmetric zero modes or,
equivalently, supersymmetric collective coordinates. 
Our main observation is that many familiar aspects of supersymmetric
instanton calculus are recovered  here from     string world-sheet Ward
identities.                             

Given the marginal character of the local I--A Hagedorn singularity
(logarithmic), it is very interesting to check the effects of
fermion zero modes at short distances. If the induced interactions
turn out to be repulsive, they could perhaps cure the instability.
In  the course of this letter we will argue in favour of the
opposite situation, namely the fermion-induced interactions
are strongly attractive at short distances.

\newsec{Collective Coordinates}
In general, there is a fermionic collective coordinate for each 
fermionic symmetry broken by the classical solution. In the Dirichlet
construction, we have to consider the space-time supersymmetries
broken by the open-string boundary conditions. 
For explicit calculations involving the supersymmetry 
 charges, it is convenient
to use the light-cone Green--Schwarz formalism. 
In the notation and conventions of \refs\rgreenf,  the bosonic
(anti-) D-instanton boundary state is a solution of the constraints
$(\alpha^i_n - \balpha^i_{-n} )|I_{\pm}, p\ket =0$, $(S^a_n \pm i           
\bS^a_{-n} ) |I_{\pm}, p\ket =0$ and may be written as the coherent
state   
\eqn\bs
{|I_{\pm}, p\ket = {\rm exp}\sum_{n=1}^{\infty}\left( {\alpha^i_{-n}
\balpha^i_{-n}\over n}
 \mp i S^a_{-n} \bS^a_{-n} \right) \, |0_{\pm}, p\ket ,}
where $S^a$ are Green--Schwarz fermions, transforming in the ${\bf 8_s}
$ of $SO(8)$, the transverse rotation group. The ground states in
\bs\ are  
the standard massless scalars  of the type-IIB string
$
|0_{\pm}, p\ket = {1\over 4}\left(|p\ket |i\ket |{\overline i}\ket
\mp i |p\ket |{\dot a}\ket |{\overline {\dot a}} \ket \right)$ 
 satisfying\foot{As usual, the indices $i,a, {\dot a}$ run in the
${\bf 8_v}, {\bf 8_s }$ and ${\bf 8_c}$ of $SO(8)$ respectively.}
 $\bra 0_{\pm}, p| = (|0_{\mp},p\ket)^{\dagger}$, $\bra
0_{+}, p |0_{+}, p' \ket = 0$, $\bra 0_{+}, p | 0_{-}, p' \ket =
\delta^{10} (p+p')  $.

There are 32  supersymmetry
 charges in the type-IIB string. From the left-moving sector we have 16
charges with $SO(8)$ quantum numbers ${\bf 8_s}\oplus {\bf 8_c}$,
 with the
 ${\bf 8_s}$ charges  given by $ 
Q^a = \sqrt{2p^+} S^a_0$, while those in the ${\bf 8_c}$ are non-linearly
realized in the light-cone gauge,  
$Q^{\dot a} = \sqrt{ 2\over \alpha'
p^+} \,\,\gamma^i_{{\dot a}a} \sum_{-\infty}^{+\infty} S^a_{-n}
\alpha^i_{n} $,  with the same structure repeated in the right-moving sector.    
 The 
combinations
$
Q_{\pm} = {1\over \sqrt{2}} (Q \pm i \bQ)$ 
 satisfy the algebra   
$
\{Q_+ , Q_- \} = \sqrt{2} \gamma_{\mu} p^{\mu} = -i\sqrt{2}
\, {\gamma\cdot\pt} $, after defining suitable $16\times 16$ gamma
matrices. The important property of the charges $Q_{\pm}$
is that they annihilate the instanton and anti-instanton:     
$Q_{\pm} |I_{\pm}, p\ket =0$. Thus, the state $|I_{\pm}\ket$ 
only breaks
the $Q_{\mp}$ supersymmetries, and we have 16 fermionic collective
coordinates in addition to the standard ten bosonic coordinates
for the position (unlike Yang--Mills instantons, these R--R gravitational
instantons are ``point-like", in the sense that no size parameter  
arises as a result of the lack of scale invariance).
     
These collective coordinates are introduced in the 
operator formalism  by the insertion of  operators:  
\eqn\coll{
e^{ix P} e^{i \theta^{\pm} Q_{\mp}}\,|I_{\pm}, x=0 \ket }  
for each instanton or anti-instanton boundary state of type \bs\ at 
fixed position. The measure for integration over
the instanton 
moduli space must be invariant under 
the unbroken   supertranslations $x\rightarrow x
+ \theta \gamma \theta $ and  is thus   determined to be   
\eqn\mea{ d\mu_{\pm} = J^{\pm} \, dx^{\pm}\, d^{16} \theta^{\pm},}
 with $J$ a convenient Jacobian, which can be obtained from the  
 analysis of the low-energy solutions (see \refs\rfint). On general
grounds, since the only bosonic zero modes are translations we know
the scaling with the string coupling $\lambda$ as $J\sim
 (\sqrt{S_{c\ell}})^{10} \sim \lambda^{-5}$, and the proportionality
constant fixes the normalizations and gives the right dimensions to
$d^{10} x_0 \, d^{16} \theta$.    

The action of $Q_{\mp}$ on $|I^{\pm} \ket$ spans two  
$2^{16}$-dimensional supersymmetry representations. In
particular, the standard fermionic zero modes correspond to the action
of one charge  
$Q^{\alpha}_{\mp} |I^{\pm}, x\ket$, which has projections (wave
functions) along any of the fermionic string states.
Clearly, the integration over the fermionic collective coordinates is
equivalent to including the complete D-instanton supermultiplet in the
path integral.
Following the rules in \refs\rpolcom, \refs\rgreengas, the 
  partition function of the instanton gas is given by 
\eqn\partfunc
{{\cal Z} = \sum_{n_+,n_- = 0}^{\infty} {1\over n_+ ! n_- !}
\prod_{j=1}^{n_+}\int d\mu_j^+ \prod_{k=1}^{n_-} \int 
d\mu_k^- \, e^{-S_{(n_+,n_-)}}, }
where the action in the $(n_+, n_-)$ instanton sector is
\eqn\action
{S_{(n_+,n_-)} = \Gamma_0 + \sum_j \Gamma_j + \sum_k \Gamma_k +
\sum_{(j_1, j_2)} \Gamma_{(j_1,j_2)} +\sum_{(k_1,k_2)}
\Gamma_{(k_1,k_2)} + \sum_{(j,k)} \Gamma_{(j,k)} + {\rm 3\,\,body 
};} 
here the index $j$ refers to instantons and $k$ to anti-instantons.
We have an expansion in irreducible many-body interaction terms, each
of them given by the sum of connected string diagrams with a number of
boundaries attached to instantons, anti-instantons, or both.
Specifically:
\eqn\genterm
{\Gamma_{(j_1,\cdots;k_1 \cdots)} = \sum_{g=0}^{\infty}
\sum_{N^+_1,\cdots =0}^{\infty} \sum_{N^-_1, \cdots =0}^{\infty}
{\lambda^{2g+\sum N^+ + \sum N^- -2} \over N^+_1 ! \cdots N^-_1 !
\cdots } W(g, N^+_j, N^-_k) .}
Here $N^{\pm}$ denote the number of boundaries attached to the same
instanton or anti-instanton, and $\lambda$ stands for 
the string coupling constant. The first term $\Gamma_0$ is the standard
perturbative sum of string diagrams {\it in vacuo}, and   
 the bare
instanton action is given by
$
\Gamma_{\pm} = {D\over \lambda}$,  
where $D$ is the disk amplitude. Diagrams with genus zero correspond
to classical interactions  between the instantons, the leading one
coming from the cylinder diagram \cilindro. The expansion \genterm\
and \partfunc\ provides a complete perturbative treatment of the
instanton interactions, including the purely classical ones,
giving a stringy version of a perturbative constrained instanton
expansion \refs\raffleck.  The BPS character
ensures that there are no classical interactions between
like-``ground-state" instantons, with boundary states given by \bs. 
In fact, a heuristic non-renormalization theorem can be
argued for the vanishing of the general string diagram with only
(anti-) instanton insertions\foot{In fact, one needs at least one
loop of closed or open strings to complete the argument, so that the
disk amplitude is non-zero even in the supersymmetric case.}.
  This is essentially Martinec's argument
for the vanishing of the cosmological constant in perturbative string
theory, using the corresponding unbroken supersymmetry. In principle,
there are
classical interactions between instanton states in the same
supermultiplet, due to the fact that instantons do interact
with anti-instantons, and both boundary states are related by the 
action of linearly realized  supersymmetry charges:
 $|0_{\pm}, p\ket = {1\over (2p^+)^4}   
\prod_{a=1}^{8} Q^a_{\pm} |0_{\mp}, p\ket$. 
Therefore, the expansion in fermionic collective coordinates 
could effectively induce bosonic I--A interactions, as in \cilindro. 
  However, we
shall see in the following that these terms do not survive the
integration over fermionic collective coordinates.   

In practice, the expressions \partfunc\ and \genterm\ are only
valid within a dilute-gas approximation, in spite of the systematic
treatment of I--A interactions involved, and  we are forced to cut-off
the integral over positions at relative separations 
 of the order of the string
scale. The Hagedorn-like instability in the I--A sector is an obvious
reason. However, even if we do not consider anti-instantons, the 
above parametrization of collective coordinates is wrong when
several D-instantons approach one another. Below the string scale,  
we have $\sim N^2$ light modes instead of the $N$ positions of a
set of $N$ D-instantons on top of each other \refs\rwittb, and
the collective dynamics corresponds to a certain $U(N)$ supersymmetric
matrix model (recently, the case $N=2$ was studied in \refs\rgreengutt). 
A very interesting possibility along these lines would be that the
effective dynamics of an I--A pair, separated by the singular distance
$\beta_{\rm Hag}/2$, is indeed described by some $U(2)$ 
matrix model.

\newsec{Ward Identity and Fermion Exchange}

It is clear that the previous setting can be used to obtain world-sheet
Ward identities associated to the supersymmetries $Q_{\pm}$. In order
to generalize the discussion, it is convenient to 
 use an abstract operator formalism in which the perturbative string
amplitude with boundary states is represented as the overlap  
\eqn\ine
{W(g, N^+_j, N^-_k) = 
\int_{\CM} \bra \Sigma_{(g, N^+_j, N^-_k)} | \Psi \ket , } 
where we integrate over the moduli space of super-Riemann surfaces
with punctures $\CM$.
The ket state has the structure 
$|\Psi \ket =  
\otimes_j  |
 x^+_j,\theta^+_j \ket \otimes_k 
 |  x^-_k,\theta^-_k \ket$ and  
 the states $|
 x^{\pm},\theta^{\pm} \ket =e^{i\theta^{\pm} Q_{\mp}} |x^{\pm} \ket$
 are one-punctured spheres,   
projected  on the instanton boundary states located
at the   point $x^{\pm}$; and  we now allow the indices $j,k$ to label
boundaries possibly located at the same space-time instanton. We 
will concentrate here on the multi-instanton vacuum amplitudes 
\genterm. However, it is clear that similar Ward identities could 
be derived for instanton-corrected scattering amplitudes, by simply
including some asymptotic scattering states in the definition of
the state $|\Psi\ket$. See \rgreengut \rhama\ for some results in this 
direction. 

The basic property we need is the contour-deformation formula to pull
an insertion of a conserved current $Q = \oint J$ 
 from one puncture to the others:
\eqn\pull
{\bra \Sigma |Q|\psi\ket \otimes_i |\chi_i\ket = \sum_i \bra \Sigma
|\psi\ket \otimes |\chi_1 \ket \otimes \cdots \otimes |(-Q)|
\chi_i \ket
\otimes \cdots }
and the  exponentiated version:  
$    \bra \Sigma |e^Q \psi \ket \otimes_i |\chi_i \ket = \bra \Sigma |\psi
\ket \otimes_i e^{-Q} |\chi_i \ket
$. 
These formal  Ward identities hold up to total derivatives in the
perturbative moduli space $\CM$, which we tacitly discard. In doing
so, our manipulations have the same heuristic status as the ``proofs"
of perturbative finiteness of superstrings.
Applying the Ward identity  to \ine\ we eliminate the fermionic
collective coordinates from one puncture, say $x^+_1$, and we get 
\eqn\pul
{\int_{\CM} \bra \Sigma |   x^+_1 \ket \otimes_{j\neq 1}  
 e^{i(\theta^+_j - \theta^+_1)Q_-} \,| 
x^+_j\ket \otimes_{k} 
e^{-i\theta^+_1 Q_-} e^{i\theta^-_k Q_+} |  x^-_k \ket .}  
Now we can use the algebra $\{Q_+, Q_- \} =-i\sqrt{2} \gamma^{\mu}
\pt_{\mu}$  
 to commute the two exponentials
acting on the anti-instanton Hilbert spaces, at the expense of
producing insertions of the momentum operator, which in turn 
translates the
anti-instanton boundary states. The $Q_-$ charge annihilates the
anti-instantons and we are left with
\eqn\otro
{\eqalign{\int_{\CM} \bra \Sigma |x^+_1 \ket &
\otimes_{j\neq 1} e^{i(\theta^+_j -
\theta^+_1) Q_-} | x^+_j -i\sqrt{2} \theta^-_1 \gamma (\theta^+_j 
-\theta^+_1)\ket
\otimes |x^-_1 + i\sqrt{2} \theta^-_1 \gamma \theta^+_1 \ket
\cr & \otimes_{k\neq 1} e^{i(\theta^-_k - \theta^-_1 )Q_+} |x^-_k +
i\sqrt{2} \theta^-_k \gamma \theta^+_1 \ket ,} }
where we have applied the same manipulations to the $x_1^-$ puncture.
In the particularly simple case of an amplitude involving only
instantons, the expression  \otro\ reduces to
$\int_{\CM} \bra \Sigma |x^+_1 \ket \otimes_{j\neq 1} |x^+_j, \theta^+_j -
\theta^+_1 \ket$, from which we can eliminate all $\theta_1$ 
dependence altogether by a  change of variables  
 in the measure: $\theta_j \rightarrow
\theta_j + \theta_1$.  
 Thus we see that the pure instanton
vacuum 
amplitudes all vanish after integration over fermionic collective
coordinates (again, up to total derivatives in the perturbative  
moduli space),
 even if, as stated before, 
 the static interactions between some states in the
supermultiplet could be non-trivial.

If $\theta^+_j = \theta^+_1 $ and
$\theta^-_k = \theta^-_1$, i.e.  when we have only one I--A  pair 
 connected by a Riemann surface, then we
have succeeded in eliminating all fermionic collective coordinates in
favour of a total derivative with respect to the relative
position. This seems
to be impossible for higher interactions (3 body, 4 body, etc). If we
truncate the instanton action to the 2-body interaction terms, as in
\action, then the partition function takes the form 
\eqn\trunca
{\eqalign{{\cal Z}_{\rm 2-body} = e^{-\Gamma_0 }
 \sum_{n_+ n_-} {1\over n_+ ! n_- !}  \int
& \prod_j dx^+_j d\theta^+_j \,J_j^+ \,
 e^{-\Gamma_j} \prod_k dx^-_k d\theta^-_k
\,J_k^- \,e^{-\Gamma_k}\times \cr & \times 
\prod_{(j,k)} {\rm exp}\left( i\sqrt{2} \theta^-_k
\gamma \theta^+_j \cdot {\partial \over \partial x_{jk}} \right)
e^{-\Gamma_{jk} (x_{jk})},} } 
where $x_{jk} = x^-_k - x^+_j $ and $\Gamma_{jk} (x_{jk})$ is given by
\iain.
Integration over the fermionic coordinates is equivalent to the
insertion of  
factors  $i\sqrt{2} \gamma^{\mu}_{\alpha\beta} \partial_{\mu} \Gamma_2
(x^- -x^+)$ for each pairing of fermionic lines, in a graphical
representation in which each instanton is an effective operator with
16 fermionic legs, labelled by the ${\bf 8_s} \oplus {\bf 8_c}$ of
$SO(8)$ Dirac indices. In eq. \cilindro, 
 $\Gamma_2 (x^- -x^+)$ 
 was interpreted as
an off-shell tree-level 
 propagator (inverse generalized Laplacian) for the
bosonic boundary states, thus   the
fermionic  
insertions are true off-shell string field theory generalizations of
fermionic propagators.
Notice that the function $
\Gamma_2 (x^- - x^+)$ may contain contributions from an arbitrary
number of boundaries and handles, so that we really obtain the
fully dressed fermionic propagators of string field theory. 

We can     also saturate the zero modes by 
applying several derivatives to
the same off-shell propagator. A term with an even number of
derivatives $(\gamma\cdot\pt)^{2n} \Gamma_2 $ can be interpreted as a
bosonic line connecting effective operators with $n$ derivatives
$\CO_b \sim \varphi^{N_b} \, \pt^n \psi^{N_f}$, with an even number of
fermions $N_f = {\rm even}$. On the other
 hand, terms with an odd number of
derivatives $(\gamma\cdot\pt)^{2n+1} \Gamma_2$ may be interpreted as a
fermionic propagator connecting fermionic operators with $n$
derivatives $\CO_f \sim \varphi^{N_b} \pt^n \psi^{N_f}$,
and  $N_f = {\rm
odd}$. 
 We thus get the stringy generalization of the
 standard rules for the  clustering
of zero modes, and the multiple-scattering approximation
 familiar from Field Theory \refs\rmottola. It is very satisfying to see
them emerge from world-sheet Ward identities in the underlying string
theory.

We can exhibit the effective operators more precisely in the     
one-instanton sector.
Fermionic zero modes pose selection rules on the effective operators
arising  in the process of ``integrating out" instanton
fluctuations. In the one-instanton sector we have to saturate the 16
zero modes of the D-instanton, and  the analogue of the resulting 
 't Hooft effective
interaction is generated to leading order in the string coupling by 16
tadpoles of fermionic vertex operators $V^{\psi}$ 
 in the instanton background.
This easily follows from the fermionic D-instanton calculus in the
form
\eqn\th
{\int dx_0\, d\theta \,J\, e^{-\Gamma_1}
 \prod_{\alpha =1}^{16} \bra V^{\psi}_{p_{\alpha}}  
|\,e^{i\theta^{\pm} Q_{\mp}} |I^{\pm}, x_0\ket_{\rm disk} =
\int dx_0 \,J\, e^{-\Gamma_1} \prod_{\alpha=1}^{16} \bra V^{\psi}_{p_{\alpha}} 
|\,Q^{\alpha}_{\mp}\,|I^{\pm}, x_0\ket_{\rm disk}. }

In general, string scattering amplitudes directly produce the
amputated Green function of string field theory. Accordingly, the
Dirichlet disk tadpoles can be interpreted as the source terms in the
classical background equations
through the formal  relation 
$$\bra V_{\Phi} (p) \,|I, x_0 \ket_{\rm disk}  \sim  (p^2 +M^2_{\Phi})
 \,
\bra \Phi (p) \ket_{I,x_0}.  
$$ 
At the massless level, this simply reduces to $\pt^{2}_x \, 
\varphi_{c\ell}(x,x_0)$, where $\varphi_{c\ell} (x,x_0)$ denotes
collectively the bosonic massless fields in the background of an
instanton located
 at $x_0$.
 These are particularly simple in
the case of the type-IIB gravitational instantons we are interested
in: at the massless level \refs\rggp,
 the D-instanton only couples to the dilaton
$\phi$ and the R--R scalar ``axion" $$i\,a_{c\ell} \sim
e^{-\phi_{c\ell}} =   
e^{-\phi_{\infty}}  \left( 1+{Q \alpha'^4 \over  
|x-x_0|^8} \right)^{-1} $$  
in the bosonic sector\foot{This  is an elementary solution
of the Laplace equation with source at $x_0$, so
$\varphi_{c\ell}(x,x_0) \sim \bra x_0 | {1\over \pt^{2}} | x\ket$ indeed
scales like a massless bosonic propagator.}, where $Q$ denotes the
R--R charge in appropriate units.  
  The fermionic zero modes ${\gamma\cdot D} 
\psi_{c\ell} =0$ appearing in \th\ are obtained from the bosonic
solution  by a
broken supersymmetry transformation. In general, starting from the
bosonic solution  
we can generate the full fermionic instanton solution as a function of
both bosonic and fermionic collective coordinates (i.e. the full
supermultiplet of instanton fields) by explicit iteration of the
type-IIB supersymmetry transformations as presented in \refs\rsch.
Formally, one finds the structure  
\eqn\stru{\eqalign{\varphi_{c\ell} (x,x_0, \theta) \sim & 
  \sum_{n=0}^{8} C_n
(\theta\, {\gamma\cdot\pt}_x \,\theta)^n \,\varphi_{c\ell} (x,x_0)   
\cr
\psi_{c\ell} (x,x_0,\theta) \sim & 
\sum_{n=0}^{7} C_n \,(\theta \,{\gamma\cdot 
\pt}_x
\,\theta)^n \,\theta \,{\gamma\cdot \pt}_x \varphi_{c\ell} (x,x_0)}} for 
bosonic and  fermionic fields respectively.  
Now, all effective operators produced by integrating out 
D-instantons are    
 generated in the local limit by the standard rule:
\eqn\rule{ \left\bra \prod_{\rm bosons} \pt^2 \varphi\, \prod_{\rm 
fermions} \gamma\cdot \pt\, \psi \right\ket_{\rm inst} \rightarrow  
\sum_{\rm local \, ops.}\bra \CO_{\rm eff}(\varphi, \psi) \ket_{\rm vac}.}
Here the factors $\pt^2$ and $\gamma\cdot\pt$ correspond to the amputation
of the Green function.
The right-hand side of \rule\ has the form    
\eqn\opss{ {\cal O}_{\rm eff} \sim e^{-D/\lambda} \,\, J\, A\, 
\varphi^{N_b} \, \pt^{N_{\pt}} \, \psi^{N_f} }
for an effective operator with $N_b$ bosons, $N_f$ fermions and
$N_{\pt}$ derivatives. Here $A$ denotes the contribution of 
non-zero modes (starting with the one-loop determinant in the
instanton background).     

More explicitly, the effective  operators are given by  
  all the different forms of  saturating the 
fermionic integration in the expression  
\eqn\efff
{ e^{-\Gamma_1} \, \int dx_0 \,d^{16} \theta \,J\,
\prod_{j=1}^{N_b}\,\pt^2_{x_j} \,
 \varphi_{c\ell} (x_j, x_0, \theta) \prod_{k=1}^{N_f}
\, {\gamma\cdot\pt}_{x_k} \psi_{c\ell} (x_k, x_0, \theta) ,}
where $\varphi (x,x_0, \theta)$ and $ \psi (x, x_0, \theta)$ are
the classical configurations in \stru,
 in complete analogy with the superfield instanton calculus as
developed for example in \ritep. 
If each bosonic leg soaks up $2n_j$
fermion collective coordinates,
 and each fermionic leg $2m_k +1$ such that $\sum
2n_j + \sum (2m_k +1) = 16$, then,  
considering the amputation factors and the
scaling of the bosonic solution $\varphi_{c\ell}\sim (\pt)^{-2}$,
 we find
a total of $\sum n_j + \sum m_k = 8- {N_f \over 2}$ derivatives in the
effective operator. In general, defining the natural dimensional index
for supersymmetric transformations $\CN \equiv N_{\pt} + N_f /2$, we have
the selection rule $\CN =  {Z \over 2}$  
 for $Z$ chiral zero modes.     

Realizing the supersymmetry in   the linearized approximation, there
 are additional selection rules coming from the BPS character of
the instantons, and the fact that the only bosonic massless couplings
correspond to the two scalars of the massless type-IIB multiplet.
The discussion is simplified by the fact that we may arrange the
massless IIB supermultiplet into a light-cone superfield containing the
antisymmetric products of the ${\bf 8_s} $ representation of $SO(8)$
 according to ref.  \refs\rgs.  
In any case, from the physical point of view, the
 upshot of this discussion of zero-mode selection rules is that
effective operators induced by ten-dimensional D-instantons of
the type-IIB theory have 16 fermions or an equivalent number of
derivatives. They are thus of very high dimension, and no interesting
infrared dynamics is induced in the one-instanton sector.  
This is true in spite of the strong Coulomb forces binding the
instantons and anti-instantons, similar to the familiar case of
2+1 gauge theories \refs\rpoly, \refs\rahw. This is consistent with
the fact that supersymmetry completely determines the type-IIB      
supergravity action to order $\CN =2$, and no potential is allowed.
On general grounds, since the instanton ``dynamics" reduces to a 
finite-dimensional
 integral, we expect the effective action after integrating
out the instanton fluctuations to be supersymmetric. This means
that  one
expects instantons to induce effective actions that could exhibit
 {\it spontaneous} breaking of supersymmetry
in some cases. On the other hand a non-supersymmetric effective action
after integrating out instantons would indicate  that supersymmetry is
anomalous. As we comment below,  a
 possible exception could be the I--A sector, because of
the Hagedorn singularity.      

On the other hand, working in less supersymmetric backgrounds, D-instantons
or more generally wrapped Euclidean D-branes could indeed generate
interesting couplings in the infrared. A notorious example appears in
\rwit, where superpotentials are generated in $N=1$ backgrounds. An
example with completely broken supersymmetry is given by the 
finite-temperature
 boundary conditions, where one readily finds periodic
potentials for the ten-dimensional axion field, and runaway potentials
for the dilaton generated by D-instantons (see \refs\rfint).

Coming back to \trunca, we can discuss the dynamical effects of
including the fermionic collective coordinates. 
 At large separations, in the dilute-gas limit, we obtain a
supressing factor $|x^+ -x^-|^{-9}$ for each paired fermion line, from
the long-distance scaling of the fermionic propagator $\gamma\cdot\pt
\,\Gamma_2 (x^+ -x^-)$. This
corresponds to the restoration of supersymmetric zero modes
(for example, factorization of determinants at one loop), and the
subsequent suppression of instanton effects.

At string-scale distances, we can estimate the fermionic propagators
in the open channel expression \iaino. Each pair of 
  fermionic collective
coordinates carries a derivative insertion and therefore an  
extra enhancement factor $|x^+-x^- -\beta_{\rm Hag}/2 |^{-1}$ in the
partition function (a logarithmic attraction term in the effective
action). This means that fermion exchange is also
 strongly attractive at
short distances.
We may consider as possible dominant configurations 
 those in which the fermionic zero modes are
saturated in ``dipole" pairs, each   one contributing  an extra factor
$|x^+ -x^- -\beta_{\rm Hag}/2 |^{-16}$. 
These configurations are also interesting because they integrate to
a total derivative in the multi I--A partition function. 
Indeed, summing 
 a dilute gas of these ``dipoles", we obtain a vacuum
 effective
action
\eqn\eff
{\Gamma_{\rm dipoles} = \Gamma_0  +  {\rm Vol}_{10}\,
\times \, e^{-2\Gamma_1}\times 2^8
 \int_{0}^{\infty}  dx\,\,{\rm
det}({\gamma\cdot \partial_x})\,\, e^{-\Gamma_2 (x)} ,}
where  the determinant is over Dirac indices in the ${\bf 8_s}\oplus
{\bf 8_c}$ of $SO(8)$.
In a normal situation, as in field theory, one would define the I--A
coincidence boundary condition at $x=0$ 
 as trivial, thereby obtaining $
\Gamma_{\rm dipoles} =0$ and unbroken supersymmetry. In this case, the
logarithmic singularity at finite separation renders \eff\ ill defined
and there is 
a possibility of supersymmetry breaking, depending on the 
boundary conditions (i.e. new physics) involving the annihilation
of the I--A pairs,  which might signal an interesting source of 
 non-perturbative
instabilities of type-IIB strings.   

In any case, it is clear that no
definite conclusions can be drawn until the physics of I--A annihilation
is understood. In fact, we can view the singularity as an effect of
not treating the extra massless modes appearing at the 
distance $\beta_{\rm
 Hag} /2$ as fully fledged collective coordinates. Therefore, one
should perhaps formulate the problem in terms of 
 the $U(2)$ matrix dynamics
of light modes of  I--A ``molecules" of approximate size $\beta_{\rm Hag} 
/2$. However, the study of
possible bound states of  the resulting I--A system is surely  complicated
by the lack of supersymmetry of these configurations.

\newsec{Conclusions}

We observe that the stringy generalization of some standard rules of
instanton calculus is simply dictated by the structure of world-sheet
Ward identities. This is the case of the handling of supersymmetry
zero modes and the corresponding fermion exchange interactions. The
consideration of anti-commuting collective coordinates does not resolve
the Hagedorn instability of overlapping I--A pairs. On the contrary, the
interaction induced by fermion exchange is strongly attractive at the
string-scale distances.

Ten-dimensional D-instantons induce effective operators with $N_{\partial}
+ {N_f \over 2} =8$, where $N_{\partial}$ is the number of derivatives
and $N_f$ is the number of fermions. Therefore, they are not very  
important in the  infrared  unless the I--A strong dynamics somehow
breaks supersymmetry. It is interesting to consider explicit D-instanton
constructions in four-dimensional $N=1$ backgrounds. In this case, in
addition to the ten-dimensional D-instantons, we would have wrapped
Euclidean D-strings, 3-branes and 5-branes, as in \rbbs. Four dimensional
superpotentials would be generated if the instanton has exactly two chiral
fermion zero modes. A very general characterization of such superpotentials
was recently put forward in \rwit. It would be interesting to work out
in more detail the weak coupling quantization of such four-dimensional
R--R instantons.

Much work is still needed in order to fully understand the 
collective dynamics of D-instantons and D-branes in general. The
occurrence of a Hagedorn singularity in non-supersymmetric sectors
may indicate that new surprises beyond the $U(N)$ enhancement
phenomenon of \rwittb\ are possible. It is then very important to
learn the semiclassical rules that reproduce non-trivial 
exact results such
as  \rvafa.

\newsec{Acknowledgements} 
It is a pleasure to thank E. Alvarez, L. Alvarez-Gaum\'e, 
C. G\'omez,  M.A. V\'azquez-Mozo, and specially
 D.J. Gross and I.R. Klebanov for useful discussions and suggestions. 

$\underline{\rm{Note}\,\,\rm{ added}\,\,\rm{in}\,\,\rm{print}}$: 
The recent preprint \refs\rgpeff\ has  some overlap with
our treatment.     

\listrefs
%\vfill\eject
\bye